\documentclass[prl,amsmath,amssymb,amsbsy,twocolumn,superscriptaddress,showpacs]{revtex4-1}
\usepackage{amsmath,amssymb}
\usepackage{wasysym}
\usepackage{graphics}
\usepackage[pdftex]{graphicx}
\usepackage{epstopdf}
\usepackage{latexsym}
\usepackage{subfigure}
\usepackage{color}
\usepackage{multirow}
\usepackage[]{natbib}
\usepackage[breaklinks,colorlinks = true,linkcolor = red,urlcolor=cyan,citecolor=red]{hyperref}
\usepackage{times}
\usepackage[abs]{overpic}

\newcommand{\be}{\begin{equation}}
\newcommand{\ee}{\end{equation}}
\newcommand{\bw}{\begin{widetext}}
\newcommand{\ew}{\end{widetext}}
\newcommand{\bea}{\begin{eqnarray}}
\newcommand{\eea}{\end{eqnarray}}
\newcommand{\ba}{\begin{array}}
\newcommand{\ea}{\end{array}}

\newcommand{\nn}{\nonumber}
\newcommand{\bs}{\boldsymbol}
\newcommand{\la}{\langle}
\newcommand{\ra}{\rangle}

\newcommand{\bk}{\mathbf{k}}

\newcommand{\hh}[1]{H^{(#1)}_Q}

\begin{document}
\title{Optical gyrotropy in quadrupolar Kondo systems}
\author{SungBin Lee}
\affiliation{Department of Physics, University of Toronto, Toronto, Ontario M5S 1A7, Canada}
\author{Arun Paramekanti}
\affiliation{Department of Physics, University of Toronto, Toronto, Ontario M5S 1A7, Canada}
\affiliation{Canadian Institute for Advanced Research, Toronto, Ontario, M5G 1Z8, Canada}
\author{Yong Baek Kim}
\affiliation{Department of Physics, University of Toronto, Toronto, Ontario M5S 1A7, Canada}
\affiliation{Canadian Institute for Advanced Research, Toronto, Ontario, M5G 1Z8, Canada}
\affiliation{School of Physics, Korea Institute for Advanced Study, Seoul 130-722, Korea}

\date{\today}
\begin{abstract}
Recent experiments point to a variety of intermetallic systems which exhibit exotic quadrupolar orders driven by the Kondo 
coupling between conduction electrons and localized quadrupolar degrees of freedom. Using a Luttinger $k \cdot p$ Hamiltonian
for the conduction electrons, we study the impact of such quadrupolar order on their energies and 
wave functions. We discover that such quadrupolar orders can induce a nontrivial Berry curvature for the conduction electron
bands, leading to a nonvanishing 
optical gyrotropic effect. We estimate the magnitude of the gyrotropic response in a candidate quadrupolar
material, PrPb$_3$, and discuss the resulting Faraday rotation in thin films.
\end{abstract}
\maketitle

Kondo coupling between conduction electrons and local quadrupolar degrees 
of freedom is of great interest for realizing the multichannel Kondo lattice model.
Candidate materials to realize this physics include Pr-based intermetallic compounds such as PrPb$_3$, Pr$T_2X_{20}$ 
(with $T$=Ir,Rh,Ti,V and $X$=Zn,Al), PrMg$_3$, PrInAg$_2$ and PrPbBi etc in which the quadrupoles
reside on $4f^2$ Pr ions.\cite{PSSB:PSSB2220980128,Galera1981317,morin1982magnetic,Giraud198541,
doi:10.1143/JPSJ.66.2566,
doi:10.1143/JPSJ.75.073705, 
PhysRevLett.106.177001,PhysRevB.86.184426,doi:10.1143/JPSJ.80.063701,
doi:10.1143/JPSJ.81.083702,doi:10.7566/JPSJ.82.043707} 
These ions possess a non-Kramers $\Gamma^3$ doublet ground state due to strong spin-orbit coupling 
and local crystal fields. Matrix elements of the dipole operator, proportional to the total angular momentum, 
vanish in this doublet Hilbert space. However, matrix elements of {\it quadrupolar} operators, rank-2 irreducible tensors
formed from the angular momentum, remain nonzero. The Doniach phase diagram suggests that strong hybridization between 
these quadrupolar doublets and the conduction electrons could lead to unusual heavy Fermi liquids, 
while weak hybridization
could lead to quadrupolar orders driven by Ruderman-Kittel-Kasuya-Yosida (RKKY) interactions.
\cite{si2010heavy,PhysRevLett.59.1240,0953-8984-8-48-012,coleman2007heavy} 

Detecting such quadrupolar orders and clarifying the nature of their
broken symmetries remain challenging issues due to a dearth of probes which couple directly to the quadrupole 
moments. 
In contrast to magnetic dipole order, the ordering of these time-reversal invariant quadrupoles does not directly manifest itself 
in nuclear magnetic resonance (NMR), muon spin rotation ($\mu$SR), or neutron diffraction measurements, necessitating
the need for indirect probes. Such probes include: (i)
ultrasonic measurements of phonon softening accompanying quadrupolar order, but this is restricted to ferroquadrupolar order;
and (ii) magnetic field induced dipolar order, which can be probed by neutron 
diffraction and whose pattern depends on the underlying quadrupolar state, but this relies on having a field regime strong
enough to induce measurable dipolar order while not significantly modifying the underlying quadrupolar order.
\cite{doi:10.1143/JPSJ.75.073705,PhysRevLett.94.197201}
 This 
experimental complexity of probing multipolar orders is also at the heart of the longstanding puzzle of ``hidden order'' in URu$_2$Si$_2$.\cite{PhysRevLett.55.2727,chandra2013hastatic}

In this Letter, 
we suggest an alternative route - the optical gyrotropic effect
\cite{landau1984electrodynamics, PhysRevB.47.11730,PhysRevB.87.115116,PhysRevB.87.165110}
- that may provide a sensitive probe of quadrupolar broken symmetries in metals.
The optical gyrotropic effect is a certain handedness in the propagation of light, leading, for instance to one circular
polarization of light propagating faster than the other. It may be observed in chiral states of
matter such as a solution of chiral glucose molecules or materials with chiral charge order. In this Letter, we argue that 
quadrupolar Kondo systems naturally provide a broad class of materials in which to expect a nonzero gyrotropic effect. The
underlying physics is simple to explain. Weak Kondo coupling of the quadrupoles to conduction electrons induces
extended RKKY interactions between them. This frustration can lead to spiral quadrupolar order, which in
turn, via the Kondo coupling, modifies the conduction electron dispersion and wave functions while preserving time-reversal
symmetry. Such
quadrupolar order breaks inversion and certain mirror symmetries, resulting in a nontrivial Berry curvature for the 
conduction electrons, and a nonzero gyrotropic response along certain high symmetry directions, measuring which can shed 
light on the nature of quadrupolar symmetry breaking.

As an illustrative example, we consider the intermetallic compound PrPb$_3$ which has been suggested to exhibit
spiral quadrupolar order below $T_Q \! \sim \! 0.4$K.\cite{onimaru2005observation} 
In PrPb$_3$, the Pr sites form a bipartite cubic lattice, so such spirals
must arise from competing further-neighbor RKKY interactions which frustrate simple ferroquadrupolar order. To study
the impact of this order on the conduction electrons, we consider the $k \cdot p$ Luttinger Hamiltonian with cubic symmetry,\cite{PhysRev.102.1030}
which describes the spin-orbit coupled Pb conduction holes near the $\Gamma$-point. Such a Luttinger Hamiltonian is
applicable to a wide variety of materials such as cubic intermetallics, GaAs, and the pyrochlore iridates.\cite{PhysRev.102.1030,PhysRevB.82.085111,PhysRevLett.111.206401} We then incorporate new 
terms allowed by the broken symmetry associated with the weak quadrupolar order in PrPb$_3$. Computing the
Berry curvature of the resulting modified band wave functions is shown to lead to a nonzero gyrotropic effect. We estimate
the magnitude of this gyrotropic response, discuss its possible signature in Faraday rotation experiments 
on thin films of PrPb$_3$, and conclude with broader implications.

\begin{figure}[t]
\scalebox{0.4}{\includegraphics{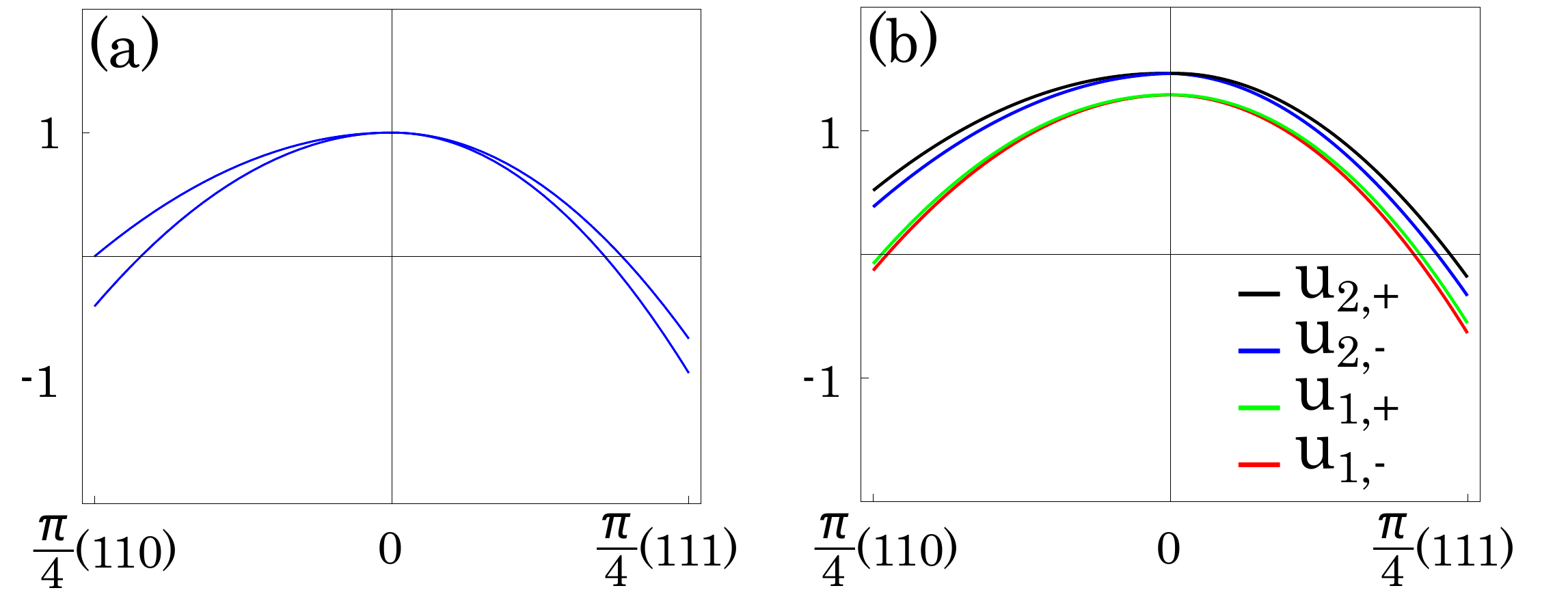}}
\caption{Band dispersion (in eV) of the Luttinger Hamiltonian in Eq.(\ref{eq:1}), with momentum measured in units of the inverse lattice
spacing. For both (a) and (b), two different momentum directions are shown: ${\bs k}$ along $(110)$ and $(111)$. 
(a) Doubly degenerate bands for the Luttinger Hamiltonian $\mathcal{H}_0$ with fitting parameters $(c_1,c_2,c_3) \!\sim\! (0.6,0.3,0.15)$eV.
(b) Band splitting due to the additional terms $\mathcal{H}_Q^{(1)}$ and $\mathcal{H}_Q^{(2)}$ with a choice
$\alpha_2/\alpha_1 \!=\! 2, \alpha_3/\alpha_1\!=\! 3$ and $ \beta_2 /\beta_1\!=\! \beta_3 /\beta_1\!=\! -1$. The overall
scale of these terms has been chosen to be significant in this figure, with $\alpha_1 \!=\! \beta_1 \!=\!0.05$eV, only in order to 
clearly depict the band splittings; for quantitative estimates of the gyrotropy in the paper, however, these are appropriately
taken to be on the scale of the quadrupolar ordering temperature $T_Q \sim 0.4$K. The eigenfunctions of $u_{1,\pm}$ and $u_{2,\pm}$ are given
in Eq.\eqref{eq:7}. }
\label{fig:1}
\end{figure}

{\it Luttinger Hamiltonian for conduction electrons. ---}
We will consider conduction electrons with spin-orbit coupling, in the presence of time-reversal and cubic crystal symmetry. 
At the $\Gamma$-point, the point group symmetry is captured by the double group of $O_h$, which
contains the maximal four-dimensional representation, the $\Gamma_8$ representation. Thus,
electronic states near the $\Gamma$-point, for instance the Pb states of PrPb$_3$, may be  described by the widely applicable four-band 
Luttinger Hamiltonian,\cite{PhysRev.102.1030} 
$H_0 = \sum_{\bk} \Psi^\dagger_{\bk\mu} h^{\mu\nu}_0(\bk) \Psi^{\vphantom\dagger}_{\bk\nu}$,
with
\bea
h_{0}(\bk) &=& - c_1 (k_x^2 \!+\! k_y^2 \!+\! k_z^2 )~ \mathbb{I}_4 - c_2 (k_x^2 J_x^2 \!+\! k_y^2 J_y^2 \!+\! k_z^2 J_z^2)  \nn \\
& &  - c_3 ( k_x k_y K_z \!+\! k_y k_z K_x \!+\! k_x k_z K_y).
\label{eq:1}
\eea
Here, $\mathbb{I}_4$ is the $4 \times 4 $ identity matrix, $J_\mu$ is the $\mu$-th component of $j=3/2$ angular momentum
matrices, and $K_\mu$ is defined via anticommutators, as $K_x = 1/2 \{ J_y, J_z \}$, $K_y = 1/2 \{ J_z , J_x \} $, and 
$K_z = 1/2 \{ J_x , J_y \}$. 
(Note that these angular momentum 
operators are distinct from the operators describing the local quadrupolar moments on Pr.) The dispersion is parameterized
in terms of unknown constants $c_{1,2,3}$.
Measuring the momentum in units of the inverse lattice constant $a$, we find that choosing $(c_1,c_2,c_3) \!\sim\! (0.6,0.3,0.15)$eV 
leads to a reasonable 
description of small hole pocket, as shown in Fig.~\ref{fig:1}(a), found near the $\Gamma$-point in {\it ab initio} band structure studies of the 
closely related compound LaPb$_3$.\cite{ram2013fermi,vijay2014}
 (We henceforth set $a=1$.) Time-reversal and inversion symmetry ensure that each band
is doubly degenerate.


{\it Quadrupolar ordering. ---} To understand the impact of quadrupole ordering
on the conduction electrons, we follow a symmetry based 
approach which considers the modification of the above Luttinger Hamiltonian $H_0$ by additional terms which are allowed by
the reduced symmetry of the quadrupolar ordered state. This enables us to directly connect the symmetry of the quadrupole
ordering with the response of the electronic states.

To illustrate this idea, let us consider quadrupole order 
with ${\bs Q}=(Q,\pi,0)$  that is known to be the ordering wave vector in PrPb$_3$  (with $Q \approx 5/4 \pi$).\cite{onimaru2005observation} This ordering breaks translational symmetry, and the original
cubic point group symmetry, but there are certain remnant symmetries: (i) ${\cal C}_{2x}$: two-fold rotations about the
$\hat{x}$-axis, (ii)
${\cal M}_{xy}$ and ${\cal M}_{xz}$: reflections in the mirror plane $xy$ or $xz$. Under these operations,
\bea
\!\!\! {\cal C}_{2x} &\! :\!& (k_y,k_z) \to (-k_y,-k_z); (J_y,J_z) \to (- J_y, -J_z)  \\
\!\!\! {\cal M}_{xz}  &:& k_y \to -k_y , ~  (J_x, J_z) \to (-J_x, -J_z) \\
\!\!\! {\cal M}_{xy} &:& k_z \to -k_z , ~ (J_x, J_y) \to (-J_x, -J_y)
\eea
Demanding that the modified Luttinger Hamiltonian be invariant under these residual symmetries, in addition to 
time-reversal symmetry, leads to
extra terms organized in powers of momentum,
\bea
\label{eq:2}
\hh1 &=& \alpha_1 J_x^2 + \alpha_2 J_y^2 + \alpha_3 J_z^2  \\
\hh2 &=& \beta_1 k_y J_z + \beta_2 k_z J_y + \beta_3 k_x (J_x J_y J_z + J_z J_y J_x).
\eea
As shown in Fig.~\ref{fig:1}(b), these terms split the two-fold band degeneracy of ${\cal H}_0$.
For a different ordering wavevector ${\bs Q}\equiv (Q_1,Q_2,0)$, 
all cubic symmetries except ${\cal M}_{xy}$ are broken, 
leading to extra terms $\beta_4 k_x J_z\!+\! \beta_5 k_z J_x\!+\! \beta_6 k_y (J_x J_y J_z \!+\! J_z J_y J_x)$;
this may be relevant to other materials.

We can compute the coefficients of these extra terms using the material-specific, symmetry allowed, Kondo couplings 
between the quadrupoles and the conduction electrons, and knowledge of the full bandstructure.
On general grounds,
since the quadrupoles are time-reversal invariant, they do not couple to the
conduction electron spin, but rather to operators such as the local density or kinetic energy.
For quadrupolar
order at wavevector ${\bs Q}$, such Kondo couplings will couple unperturbed electronic states at momenta
$(\bk,\bk+\bs{Q})$. Assuming that these states differ in energy by a characteristic energy scale $\Delta \varepsilon$,
second order perturbation theory suggests that $\alpha_i, \beta_i \sim V^2/\Delta\varepsilon$,
where $V$ is the strength of the Kondo coupling. More physically, since these terms arise below the quadrupolar transition
temperature $T_Q$, which in turn is determined by the RKKY coupling $\sim V^2/\Delta\varepsilon$, we expect $\alpha_i, \beta_i \sim T_Q$.

We next study the impact of $\hh1,\hh2$ on the optical gyrotropy
of the quadrupolar ordered state. Denoting the energy scale of the $k \cdot p$ Hamiltonian by $W$,  two distinct
momentum regimes emerge naturally: (i) small momentum, where
$k \ll T_Q/W$, and (ii) large momentum, where $k \gg T_Q/W$.
We next turn to an analytical perturbative approach to compute the Berry 
curvature and gyrotropic response arising from these regions of momentum space.

{\emph{Small momentum, $k \ll T_Q/W$. ---}
In this limit, we can start at the $\Gamma$-point where $\hh1$ breaks the four-fold degeneracy of the unperturbed
states, leading to a pair of Kramers doublets. The leading corrections away from the $\Gamma$-point arise from 
linear-in-momentum terms present in $\hh2$, which weakly splits these Kramers pairs. 

For the $(Q,\pi,0)$ order relevant to PrPb$_3$, projecting
to the eigenstates of $H_0 + \hh1$ at the $\Gamma$-point, only terms
$\beta_1,\beta_2$ in $\hh2$ have nonzero matrix elements, while the matrix element of the $\beta_3$ term
vanishes. The splitting of each doublet away from the $\Gamma$-point can then be described in terms 
of just two Pauli matrices, leading to a vanishing Berry curvature over any small momentum patch. Along 
$k_y=k_z=0$, the doublet remains unsplit, leading to a line node, and the Berry curvature is not well-defined
on a patch which intersects this line node. However, the gyrotropic response, as discussed below, 
integrates the Berry curvature over all occupied bands; thus, patches near the line node also do not contribute 
to the final result since the contributions from the two bands touching at the line node will mutually cancel.
Thus, we expect the small momentum region gives a vanishing contribution to the gyrotropy in PrPb$_3$.
Note, however, that for more general $(Q_1,Q_2,0)$ order, this is no longer true.

{\emph{Large momentum, $k \gg T_Q/W$. ---}
In this limit, we first diagonalize $\mathcal{H}_0$ (see Supplemental Material for energies and eigenfunctions) 
which leads to two pairs of doubly degenerate bands, and
project $\hh1$ and $\hh2$ into each degenerate manifold.
The projection of $\hh1$ turns out to be proportional to the identity matrix, and hence does not modify the
eigenstates which remain degenerate. We thus focus on the effects of $\hh2$, and express its
projection into each
doublet of $\mathcal{H}_{0}$ using Pauli matrices, as
$\tilde{\mathcal{H}} _{\ell} = \vec f_\ell ({\bs k}) \cdot \vec \sigma$,
where $\ell=1,2$ labels the two different degenerate bands. The eigenvalues are
given by $E_{\ell, \pm} (\bk) = \epsilon_\ell (\bk) \pm | \vec f_\ell (\bk)|$,  where $\epsilon_\ell ({\bs k})$ are the
unperturbed (degenerate) band energies, and the eigenfunctions are
\bea
u_{\ell,\pm}(\bk) = 
\begin{pmatrix} \pm \sqrt{ \frac{1}{2} \pm \frac{f_{\ell,x(\bk)}}{2| \vec f_\ell (\bk)| } } \\
\frac{f_{\ell,x}(\bk) + i f_{\ell,y} (\bk)}{\sqrt{2 (|\vec f_\ell(\bk)|^2 \pm f_{\ell,z}(\bk)  |\vec f_{\ell}(\bk)|)}}
 \end{pmatrix}.
\label{eq:7}
\eea
Fig.~\ref{fig:1} (b) illustrates how doubly degenerate bands in Fig.\ref{fig:1} (a) split into four bands 
with  eigenfunctions $u_{\ell ,\pm} ({\bs k})$ 
in the presence of ${\mathcal{H}}_Q^{(2)}$, for given parameters $\alpha_2/\alpha_1 =2, \alpha_3 /\alpha_1 =3$ 
and $\beta_2 / \beta_1 = \beta_3/\beta_1 =-1$. 
The overall
scale of these terms has been chosen to be significant in Fig.~\ref{fig:1}(b), with $\alpha_1 \!=\! \beta_1 \!=\!0.05$eV, only in order to 
clearly depict the band splittings; however, for quantitative estimates of the gyrotropy discussed below, these are correctly
taken to be on the scale of $T_Q \sim 0.4$K.
For given eigenfunctions say $u_{\ell,\pm}$, the $\gamma$ component of 
Berry curvatures $\Omega_{\ell,\pm}^\gamma$ are defined as
\bea
\Omega_{\ell,\pm}^\gamma = i \epsilon_{\mu \nu \gamma}
 \la \partial_{k_\mu} u_{\ell,\pm } | \partial_{k_{\nu}} u_{\ell,\pm } \ra, 
\eea
where $\epsilon_{\mu \nu \gamma}$ is the totally antisymmetric tensor.
A nonzero Berry curvature $\Omega_{\ell,\pm}^\gamma$
induces an anomalous velocity of conduction electrons parallel to $\mu$ direction 
when the electric field is applied along $\nu$ direction. 
In the time reversal invariant but lattice symmetry broken system, 
such anomalous velocity induces the transverse current $j_\mu = \lambda_{\mu \nu \gamma}^G d E_\nu / d x_\gamma$ 
that is proportional to the gyrotropic coefficient $\lambda_{\mu \nu \gamma}^G$. 
In particular, for oscillating electric field $E_\nu (x_\gamma ,t) = \text{Re} [ E e^{i (q x_\gamma - \omega t )}]$ 
that propagates along the ${\gamma}$ direction with wave vector $q$ and frequency $\omega$, 
the gyrotropic coefficient $\lambda_{\mu \nu \gamma}^G$ can be derived within the relaxation time approximation,\cite{PhysRevB.87.165110} 
\bea
\lambda_{\mu \nu \gamma}^G \approx \frac{e^2}{\hbar} \frac{2}{ (2\pi)^3} \frac{l_{mf}}{ (1- i \omega \tau)^2} 
\int_{k_\gamma >0} d k_\gamma \Phi^\gamma (k_\gamma), 
\label{eq:5}
\eea
for $q l_{mf} \ll 1$ 
where $l_{mf} = v_F \tau$ is a mean free path with Fermi velocity $v_F$ and relaxation time $\tau$, 
and $ \Phi^\gamma (k_\gamma) \equiv \int_{occ} d k_{\mu }d k_{\nu}~ \Omega^\gamma  ({\bs k})$ is 
the 2D integral of Berry curvatures for a given $k_\gamma >0 $.
The detailed derivation of $\lambda_{\mu \nu \gamma}^G$ in Eq.\eqref{eq:5} 
is shown in the Supplemental Material.

We begin by discussing how symmetry constrains the gyrotropic effect.
The $\gamma$ component of Berry curvature is odd under the mirror symmetries 
if a mirror plane contains $\gamma$ direction. 
Therefore, for quadrupole order with the wave vector ${\bs Q} = (Q,\pi,0)$, 
since mirror symmetries ${\cal M}_{xy}$ and ${\cal M}_{xz}$ are present,
this guarantees $\Phi^x (k_x) = \Phi^y (k_y) = \Phi^z (k_z)=0$. Thus, we expect
$\lambda_{\mu \nu \gamma}^G = 0$ for $\gamma= [100], [110]$ directions.
However, there are no mirror planes for $\gamma=[111]$, so we expect $\Phi^{\gamma} (k_\gamma) \neq 0$ for this
direction.

To estimate the gyrotropy for $\gamma=[111]$, we compute the 2D integration of Berry curvatures $\Phi^\gamma (k_\gamma)$ 
including all four bands, based on the eigenvalues and eigenfunctions of $\tilde{\mathcal{H}}_\ell$. 
For a spherical hole pocket, one can further simplify the integration into only angle dependence, 
realizing that $\int d k_\gamma \Phi^\gamma (k_\gamma)$ does not depend on the Fermi wave vector in the limit of 
 $k  \gg T_Q /W $. 
Now, we consider the surface integration of $\Phi^\gamma (k_\gamma)$ in Eq.\eqref{eq:5}
 \bea
 \int_{k_\gamma >0} d k_\gamma \Phi^\gamma (k_\gamma) 
 \approx  \frac{1}{W} 
 \int_{\theta,\varphi}  ( | \vec{ f}_1 |~\Omega_{1,+}^\gamma  +|\vec{ f}_2|  ~\Omega_{2,+}^\gamma) . 
 \label{eq:10}
 \eea
Here, the integration of azimuthal and polar angles $\varphi$ and $\theta$ for upper half of a sphere
is $\int_{\theta,\varphi} = \int_0^{\pi/2} \sin{\theta} d \theta \int _0^{2\pi} d \varphi $, 
where ${\bs k}$ is defined as $(k_\mu, k_\nu, k _\gamma) = k( \sin{\theta} \cos{\varphi}, \sin{\theta} \sin{\varphi}, \cos{\theta})$ 
in spherical coordinates. (See the Supplemental Material for a detailed derivation of Eq.\eqref{eq:10}.)

Fig.~\ref{fig:2} (a) and (b) show the azimuthal and polar angle dependence of $ | \vec{f}_1|~\Omega_{1,+}^\gamma $ 
and $ | \vec{f}_2 |~\Omega_{2,+}^\gamma$ (in units of $T_Q$) respectively for 
$c_2=0.3$eV, $c_3 =0.15$eV (results do not depend on $c_1$) and $\beta_1 = - \beta_2 = - \beta_3=T_Q$. 
One can easily confirm that $ | \vec{f}_\ell |~\Omega_{\ell,+}^\gamma $ 
is odd under time reversal symmetry 
: $\theta \rightarrow \pi - \theta$ and $\varphi \rightarrow \pi + \varphi$. 
Integration of upper half of a sphere for both $ | \vec{f}_\ell |~\Omega_{\ell,+}^\gamma $ 
results in 
$ \int_{\theta,\varphi}  ( | \vec{f}_1 |~\Omega_{1,+}^\gamma  + |\vec{f}_2| ~\Omega_{2,+}^\gamma)  = \eta T_Q $,
with the coefficient $\eta \sim 0.1$-$1$
depending on the relative signs of $\beta_1, \beta_2$ and $\beta_3$. 


\begin{figure}[t]
\scalebox{0.25}{\includegraphics{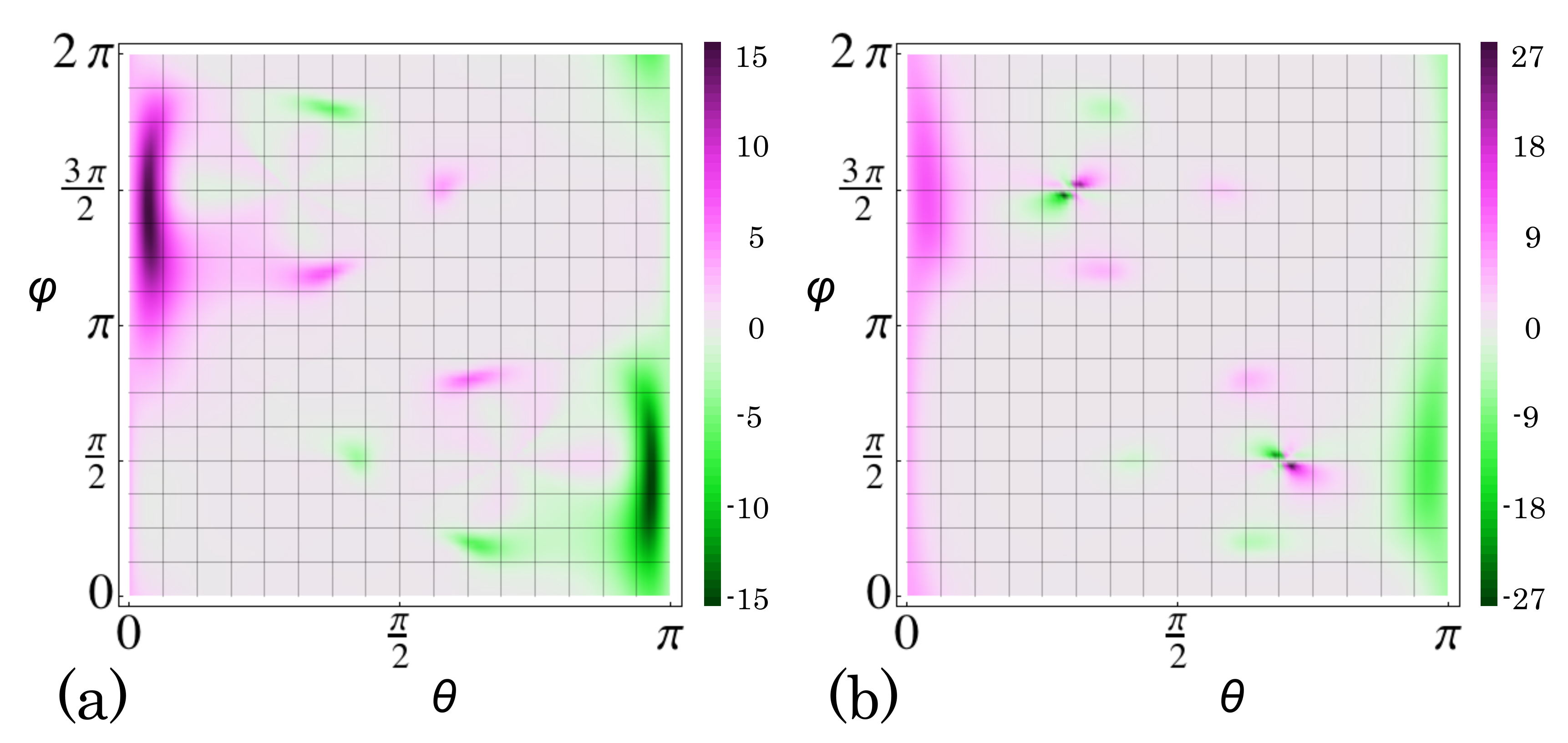}}
\caption{Azimuthal and polar angle dependence of $ |\vec{f}_1| ~\Omega_{1,+}^\gamma $ 
and $ |\vec{f}_2| ~\Omega_{2,+}^\gamma $ in units of $T_Q$, with given parameters 
 $c_2 = 0.3$eV, $c_3=0.15$eV (these results are independent of $c_1$) and $\beta_1 = - \beta_2 = - \beta_3=T_Q$ when $\gamma =[111]$. 
See the main text for details.  }
\label{fig:2}
\end{figure}


\emph{Experimental Signature. ---}
So far, we have considered 
how quadrupole order can induce band Berry curvature and a nonzero gyrotropic coefficient $\lambda^G_{\mu\nu\gamma}$.
Such gyrotropy will lead to a finite Faraday rotation of light transmitted through thin films with quadrupolar order.
Including the gyrotropic coefficient $\lambda_{\mu \nu \gamma}^G$, 
the conductivity tensor can be represented as 
$\sigma_{\mu \nu } (\omega , k) \simeq \sigma(\omega) \delta_{\mu \nu} + i \lambda_{\mu \nu \gamma}^G k_{\gamma} $.
Using Maxwell equations $\nabla \times {\bs B} = \mu_0 {\bs j} $ 
and $\nabla \times {\bs E} = - \frac{\partial{\bs B}}{\partial t}$ with vacuum permeability $\mu_0$, the complex index of refraction $N \equiv ck/\omega$ ($c$ is the speed of light) can be written up to the leading order of $\lambda^G_{\mu \nu \gamma}$ : $N_\pm \approx (n + i \kappa) \pm {  i \mu_0 c \lambda^G_{\mu \nu \gamma}}/{2}$ for right- or left-circularly polarized electric fields ${\bs E} = ( \hat{x}_\mu \pm  i \hat{x}_\nu ) E_0 e^{ i (k x_\gamma - \omega t)} $, where $n+i \kappa$ is the normal complex refractive index in the absence of gyrotropy.
This difference of refractive indices leads to the rotation of the principal axis polarization of incident light,
with the Faraday rotation angle $\theta_F$ given by
\bea
\theta_F 
&=& \frac{\omega d}{ 2c} \text{Re} [ N_+ - N_-] 
\approx  \frac{\omega d}{2c} \text{Im} [ \mu_0 c \lambda^G_{\mu \nu \gamma}] \nn \\
& \approx &  \frac{\omega d}{2c}  \frac{\alpha  l_{mf} }{\pi^2} \text{Im} [\frac{1}{(1- i \omega \tau)^2}] 
\int_{k_{\gamma}>0} d k_{\gamma} \Phi^\gamma ( k_{\gamma}) , \nn
\eea
where we have used Eq.\eqref{eq:5} for the gyrotropic coefficient $\lambda^G_{\mu\nu\gamma}$.
Here, $d$ is the film thickness of materials, $\alpha = e^2 / (4\pi \epsilon_0 \hbar c)$ is the fine structure constant,
and $l_{mf}=v_F \tau$ is the mean free path determined by the Fermi velocity $v_F$ and the transport lifetime $\tau$.
Next, let us estimate the Faraday rotation angle per unit thickness for PrPb$_3$ films.
To maximize $\theta_F/d$, we need to work at frequencies $\omega \sim 1/\tau$, which leads to
\be
\frac{\theta_F}{d} \sim \frac{\alpha}{2\pi^2} \frac{v_F}{c} \int_{k_{\gamma}>0} d k_{\gamma} \Phi^\gamma ( k_{\gamma}),
\ee
the estimate of which depends the Fermi velocity $v_F$ and the integrated Berry curvature.

{\emph{(1) Fermi velocity and transport lifetime. ---} 
The electronic bands have been studied for LaPb$_3$ where La$^{3+}$ forms closed shell and is inert 
without quadrupolar order from Pr$^{3+}$ $4f^2$ non-Kramers doublet.
It shows the hole pockets near $\Gamma$ point in LaPb$_3$ due to the conduction electrons of Pb, 
which have been also observed for PrPb$_3$ in de Haas-van Alphen experiments.\cite{ram2013fermi, vijay2014,aoki1997fermi}  
Strong spin-orbit coupling of Pb conduction electrons suggests that the Pb electronic states are well described by the Luttinger 
Hamiltonian ${\cal H}_0$ in Eq.\eqref{eq:1}. Based on the band structure calculation of LaPb$_3$, 
we estimate the Fermi velocity $v_F \approx 2 \times 10^6 m/s$.

Furthermore, we obtain the carrier density $n_d = 4 \times k_F^3/(6\pi^2) \approx 2.5 \times 10^{26} m^{-3}$ arising from the
four hole pocket bands with $k_F a \approx \pi /4$ (lattice constant $a \sim 5 \AA$),  with an effective mass $m^* \approx 0.1 m_e$. 
Within a Drude picture, the complex conductivity is written as 
$\sigma (\omega) = \sigma_0 /(1- i \omega \tau)$, where $\sigma_0 = n_d e^2 \tau/m^* $.
Close to the quadrupole ordering temperature $T_Q$,  $\sigma_0 \approx 1 \mu \Omega^{-1} cm^{-1}$ for PrPb$_3$, which
yields $\tau \approx 10^{-12} s$. We thus need to probe the gyrotropy at a frequency $\omega \approx 10^{12}$Hz.

{\emph{(2) Integrated Berry curvature. --- } 
As mentioned earlier, PrPb$_3$ exhibits spiral quadrupolar order at $T_Q \approx 0.4$K with an ordering wave vector 
${\bs Q} \approx (Q,\pi,0)$ where $Q \approx 5/4 \pi$. Based on our previous discussion,
the integrated Berry curvature from the reconstructed bands is 
$\int_{k_{\gamma}>0} d k_{\gamma} \Phi^\gamma ( k_{\gamma}) \approx \eta T_Q/W$ (with $\eta \sim 0.1$-$1$).
This leads to the final result
\be
\frac{\theta_F}{d} \sim \frac{\alpha}{2\pi^2 a} \left(\frac{v_F}{c} \right) \left(\frac{\eta T_Q}{W} \right).
\ee
where we have reinstated the lattice constant $a$.

{\emph{(3) Faraday angle. --- } 
Since PrPb$_3$ is metallic, we need to estimate the optimal film thickness for which we expect significant transmission of light, with
the maximum possible Faraday angle.
The penetration depth $\delta$ of light is determined by the complex refractive index $N= n + i \kappa$ via
$\delta = c / \omega  \kappa$. 
For $\omega \approx 10^{12}$Hz (i.e., infrared light), and a complex Drude conductivity with $\sigma(\omega) = \sigma'(\omega) + i \sigma''(\omega)$,
we find $\kappa \approx \sqrt{ \mu_0  c^2 \sigma '(\omega) / \omega} \approx 3 \times  10^{3} $, so the estimated penetration depth is 
$\delta = c/ \omega \kappa \approx 100 ~nm $, and we must choose the film thickness $d \sim \delta$.

Thus, for $T < T_Q$, when the infrared light with a frequency $\omega \approx 10^{12} Hz$ is transmitted through PrPb$_3$ 
along the $[111]$ direction, setting $\eta \gtrsim 0.1$ yields
a Faraday rotation angle $\theta_F \gtrsim 0.01 \mu$rad
for a film thickness $ d \approx 100~nm$. 
Such small rotation angles have been measured in Kerr experiments on superconducting materials at very low temperatures, 
so we expect the Faraday rotation to be accessible in future experiments.
\cite{PhysRevLett.97.167002,PhysRevLett.100.127002}

{\it Conclusion. ---} 
Motivated by a variety of intermetallic systems that exhibit quadrupolar orders, 
we have proposed that optical gyrotropy in quadrupolar Kondo systems
\cite{morin1982magnetic,  Giraud198541, PhysRevLett.106.177001, PhysRevB.86.184426, 
doi:10.1143/JPSJ.80.063701, doi:10.1143/JPSJ.81.083702, PhysRevLett.87.057201}
may yield further information on the nature of broken symmetry.
Kondo coupling between conduction electrons and localized quadrupolar moments 
can naturally drive interesting quadrupolar order and further induce non-trivial Berry curvature for conduction electrons. 
Such effect leads to a certain handedness of light propagation, resulting in Faraday rotation. 
Using a Luttinger $k \cdot p$ Hamiltonian for conduction electrons, we explored how quadrupolar order impacts the
gyrotropy. Finally, we 
estimated the Faraday rotation angle for a candidate material PrPb$_3$, which can be measured in future experiments. 
The optical gyrotropic effect might thus serve to shed further light on ``hidden" quadrupolar orders. \\

We are grateful to S. Bhattacharjee, V. S. Venkataraman and J. Orenstein for useful discussions. 
We acknowledge support from NSERC of Canada (SBL, YBK, AP) and
the Center for Quantum Materials at University of Toronto (SBL, YBK).

\bibliography{gyrotropy-refs}

\section{Supplemental material}
\label{sec:supp}

\subsection{Eigenfunctions of the Luttinger Hamiltonian and their perturbations }
\label{subsec:luttinger }

We first write the expression of energies and eigenfunctions of the Luttinger $k \cdot p$ Hamiltonian $\mathcal{H}_0$ 
presented in Eq.\eqref{eq:1} 
and then describe the projection of $\mathcal{H}_{Q}^{(2)}$ into those eigenfunctions.

The two doublet eigenstates $(\psi_1, \psi_2, \psi_3, \psi_4)$ and their energies 
$(\epsilon_1 , \epsilon_1,\epsilon_2 , \epsilon_2)$ of Eq.\eqref{eq:1} are described by,
\begin{widetext}
\bea
\epsilon_{1(2)}   &=& - \frac{1}{4} \Big( 
(4 c_1 + 5 c_2) (k_x^2 + k_y^2 + k_z^2 ) \pm 
2 \sqrt{ 4 c_2^2~ (k_x^4 + k_y^4 + k_z^4 - k_y^2 k_z^2 - k_x^2 k_z^2 -k_x^2 k_y^2 )  
+ 3 c_3^2~ ( k_y^2  k_z^2 + k_x^2 k_z^2 + k_x^2 k_y^2 ) }
\Big), \nn  \\ \\
\psi_1 &=& \frac{1}{N} \left(
\begin{array}{c}
a_1 - i a_2   \\
a_3 + a_4  \\
0 \\
a_5 + i a_6
\end{array}
\right), ~ ~
\psi_2 =\frac{1}{N} \left(
\begin{array}{c}
-a_5 + i a_6 \\
0 \\
-a_3 - a_4 \\
a_1 + i a_2 
\end{array}
\right), ~ ~
\psi_3 = \frac{1}{N} \left(
\begin{array}{c}
0  \\
a_5 - i a_6 \\
-a_1 + i a_2 \\
-a_3 - a_4
\end{array}
\right), ~ ~
\psi_4 =\frac{1}{N} \left(
\begin{array}{c}
a_3 + a_4  \\
- a_1 - i a_2  \\
-a_5 - i a_6 \\
0
\end{array}
\right) , \\
a_1 &=& 2 \sqrt{3} c_3 k_x k_z,~~ a_2 = 2 \sqrt{3} c_3 k_y k_z, ~~
a_3 =  c_2 (k_x^2 + k_y^2 - 2 k_z^2 ),   \\
a_4 &=& 2 \sqrt{ 4 c_2^2~ (k_x^4 + k_y^4 + k_z^4 - k_y^2 k_z^2 - k_x^2 k_z^2 -k_x^2 k_y^2 )  
+ 3 c_3^2~ ( k_y^2  k_z^2 + k_x^2 k_z^2 + k_x^2 k_y^2 ) },  \\
a_5 &=&  2 \sqrt{3} c_2 ( k_x^2 - k_y^2 ), ~~ a_6 =  2 \sqrt{3} c_3 k_x k_y,  \\
N &=&  \sqrt{ a_1^2 + a_2^2 + (a_3 + a_4 )^2 + a_5^2 +a_6^2 }.
\eea
\end{widetext}

Using eigenfunctinos $\psi_1,\psi_2,\psi_3$ and $\psi_4$, 
one can derive the projection of $\mathcal{H}_{Q}^{(2)}$ into those eigenfunctions,
\bea
\tilde{\mathcal{H}}_{1} &=& \la \psi_{1(2)} | \mathcal{H}_{Q}^{(2)} | \psi_{1(2)} \ra 
=  \vec f_1 ({\bs k}) \cdot \vec \sigma 
\label{eq:20} \\
\tilde{\mathcal{H}}_{2} &=& \la \psi_{3(4)} | \mathcal{H}_{Q}^{(2)} | \psi_{3(4)} \ra 
=  \vec f_2 ({\bs k}) \cdot \vec \sigma . 
\label{eq:21}
\eea

\subsection{Gyrotropic effect}
\label{subsec:gyrotropy}

The gyrotropic coefficient is related to a transverse current proportional to the gradient of electric field: 
$j_\mu = \lambda_{\mu \nu \gamma}^G d E_{\nu} /d x_\gamma$. Here, we consider an applied electric field 
$E_\nu (x_\gamma ,t) = \text{Re}[ E e^{ i (q x_\gamma - \omega t )}]$ 
and derive the expression of gyrotropic coefficient $\lambda_{\mu \nu \gamma}^G$. We note that this derivation 
is already studied in Ref.\onlinecite{PhysRevB.87.165110}.
For a finite Berry curvature, the velocity of an electron wave packet with a band dispersion $\epsilon({\bs k})$ is given by, 
\bea
{\bs v} ({\bs k}) = \frac{1}{\hbar} \frac{ \partial \epsilon ({\bs k}) }{\partial {\bs k}} - \frac{e}{\hbar} {\bs E} \times \Omega ({\bs k}).
\label{eq:22}
\eea
Based on Eq.\eqref{eq:22}, the local transverse current carried by electrons whose wave vectors lie in a slice of $k$-space of thickness $d k_\gamma$ is described by
\bea
J_{loc} ( k_\gamma ,x_\gamma,t) &=&  - \frac{e^2}{\hbar} \frac{ d k_\gamma}{4\pi^2} E e^{i ({ q x_\gamma - \omega t})} \Phi (k_\gamma)  \\
 \Phi^\gamma (k_\gamma) &=& \int_{occ} d k_{\mu }d k_{\nu}~ \Omega^\gamma  ({\bs k}).
\label{eq:14}
\eea 
The transverse {\sl nonlocal} current, $J_{nl} (x_\gamma,t)$ is the summation of local current $J_{loc} (k_\gamma, x_\gamma',t)$ for $x_\gamma>x_\gamma'$ and $x_\gamma' > x_\gamma $
\begin{widetext} 

\bea
J_{nl} (k_\gamma ,x_\gamma,t) &=& \frac{1}{ l_{mf}}  \int d x_\gamma' ( J_{loc}^{x_\gamma'> x_\gamma} 
+ J_{loc}^{x_\gamma > x_\gamma'})  \\
\int d x_\gamma' J_{loc}^{x_\gamma'>x_\gamma}  &=&  \int_{x_\gamma}^\infty d x_\gamma' ~ p (x_\gamma- x_\gamma') J_{loc} (k_\gamma ,x_\gamma' , t'(x_\gamma') )    \nn \\ 
&=& - \frac{e^2}{{\hbar}} \frac{d k_\gamma}{4 \pi^2} ~  \Phi (k_\gamma) ~E
  \int_{x_\gamma}^\infty d x_\gamma'  ~ \text{exp} [ -{ \frac{ (x_\gamma'-x_\gamma) }{ l_{mf}} } ] ~ 
  \text{exp} [ {  i q x_\gamma' - i \omega ( t - \frac{ ( x_\gamma'-x_\gamma) }{ v_F} )} ]  
  \nn \\
&=& - \frac{e^2}{\hbar} \frac{d k_\gamma}{4 \pi^2} ~  \Phi (k_\gamma) ~E
\int^{\infty}_0 d \tilde{x}_\gamma ~  \text{exp} [ - \frac{ \tilde{x}_\gamma } {l_{mf} } + i q (\tilde{x}_\gamma + x_\gamma) - i \omega t 
+ \frac{i  \omega {\tilde{x}_\gamma }  }{ v_F} ] \nn \\
&=& -  \frac{e^2}{\hbar} \frac{d k_\gamma}{4 \pi^2} ~ \Phi (k_\gamma)  ~ E e^{i (q x_\gamma - \omega t)} 
 \frac{ l_{mf} }{1 - i \omega \tau - i q l_{mf}}  \\
\int d x_\gamma' J_{loc}^{x_\gamma > x_\gamma'} &=& - \frac{e^2}{\hbar} \frac{d k_\gamma}{4 \pi^2} ~  \Phi (k_\gamma) ~E
  \int_{- \infty}^{x_\gamma} d x_\gamma'~ \text{exp} [- \frac{ (x_\gamma-x_\gamma')}{ l_{mf}} ]~ \text{exp} [ i q x_\gamma' - i \omega (t - \frac{(x_\gamma -x_\gamma')}{ v_F} ) ] \nn \\
&=&   - \frac{e^2}{\hbar} \frac{d k_\gamma}{4 \pi^2}~  \Phi (k_\gamma) ~E
\int_0^\infty  d \tilde{x}_\gamma ~  \text{exp} [-  \frac{ \tilde{ x}_\gamma }{ l_{mf}}  - i q (\tilde{x}_\gamma - x_\gamma) - i \omega t + \frac{ i \omega \tilde{x}_\gamma }{ v_F} ] \nn \\
&=&  - \frac{e^2}{\hbar} \frac{d k_\gamma}{4 \pi^2} ~ \Phi (k_\gamma)  ~ E e^{i (q x_\gamma - \omega t)} 
 \frac{l_{mf} }{1 - i \omega \tau + i q  l_{mf}}   \\
 J_{nl} (x_\gamma,t) &=&  - \frac{e^2}{\hbar} \frac{1}{(2\pi)^3} E e^{i (q x_\gamma -\omega t)}
 \Big( \int_{k_\gamma < 0 } dk_\gamma~ \frac{ \Phi (k_\gamma)}{ 1- i \omega \tau - i q l_{mf} } +
 \int_{k_\gamma > 0 } dk_\gamma~ \frac{\Phi (k_\gamma)}{ 1- i \omega \tau + i q l_{mf} }  \Big) \nn \\
 &\approx&
 \frac{e^2}{\hbar} \frac{2}{(2  \pi)^3}   ~ E e^{i (q x_\gamma - \omega t)} \int_{k_\gamma >0} d k_\gamma ~ \Phi (k_\gamma)
 \frac{ i q  l_{mf}}{ (1- i \omega \tau)^2}, ~\text{ for } q~ l_{mf} \ll 1 \nn \\
 J_{nl} (x_\gamma,t) &=&  \lambda_{\mu \nu \gamma}^G ~ i q ~ E e^{ i( q x_\gamma -\omega t)}  \\
 \lambda_{\mu \nu \gamma}^G &=&  \frac{e^2}{\hbar} \frac{2}{(2 \pi)^3} \frac{ l_{mf} }{ (1 - i \omega \tau)^2 } 
 ~ \int_{k_\gamma >0}  d k_\gamma  \Phi (k_\gamma),~ ~\text{ for } q ~ l_{mf} \ll 1
 \label{eq:23}
\eea
\end{widetext}
within the relaxation time approximation $ p(x_\gamma- x_\gamma') \propto e^{ - |x_\gamma- x_\gamma'| / l_{mf}}$ with a mean free path $l_{mf} = v_F \tau$.

Now, we are interested in the derivation of Eq.\eqref{eq:10} within the spherical Fermi surface approximation. 
The spherical coordinates are defined using radial, azimuthal and polar coordinates $ (k,\varphi,\theta)$; 
${\bs k} = ( k_\mu , k_\nu , k_\gamma ) = 
( k \sin {\theta} \cos{\varphi} , k \sin{\theta} \sin{\varphi}, k \cos{\theta} )
$
\begin{widetext}
\bea
\int_{k_z >0}  d k_z ~\Phi (k_z)  &=& \int_0^{\infty}  k^2 dk  \int_0^{\pi/2} \sin {\theta}~ d {\theta} \int_0^{2 \pi} d \varphi ~
\Omega_{\text{occ}}^{\gamma} ( k ,\theta,\varphi) \nn \\
& =  &\int_0^{\infty} k^2 d k   \int_0^{\pi/2} \sin {\theta} ~d {\theta} \int_0^{2 \pi} d \varphi ~
k^{-2} \sum_{i=1}^2 \sum_{j=\pm}   \Omega_{i,j}^{\gamma} (\theta,\varphi) n_f ( E_{i,j} ( k, \theta, \varphi)) \nn \\
  & \approx & 
- 2  \int_0^{\infty}  d k   \int_0^{\pi/2} \sin {\theta} ~ d {\theta} \int_0^{2 \pi} d \varphi ~
 ~  \Big(  \Omega_{1,+}^\gamma (  \theta , \varphi ) ~ | \vec{ f} _1 (k, \theta, \varphi) |  ~  \delta ( \epsilon_1 (k, \theta, \varphi) - \mu)  \nn \\
 && ~~~~~~~~~~~~~~~~~~~~~~~~~~~~~~~~~~~~~~~
+  \Omega_{2,+}^\gamma (  \theta , \varphi  ) ~  |\vec{f}_2 (k, \theta, \varphi) | ~  \delta ( \epsilon_2 (k, \theta, \varphi) - \mu)  \Big)  \nn \\
& \approx & - \frac{1}{a}  \frac{1}{W}
 \int_0^{\pi/2} \sin {\theta} ~ d {\theta} \int_0^{2 \pi} d \varphi ~
 ~  \Big(  \Omega_{1,+}^\gamma (   \theta , \varphi ) ~  | \vec{f}_1 ( \theta, \varphi) |
 + \Omega_{2,+}^\gamma (  \theta, \varphi ) ~  | \vec{f}_2 (\theta, \varphi) | \Big)
 \label{eq:24}
\eea
\end{widetext} 
where $n_f (E) $ is the Fermi distribution with energy $E$, $a$ is a lattice constant, four distinct energies are $E_{\ell,\pm} ({\bs k}) = \epsilon_\ell ({\bs k}) \pm  |\vec{f}_{\ell } ({\bs k} )|  $. 
($\vec{f}_\ell ({\bs k}) $ is defined in Eqs.\eqref{eq:20} and \eqref{eq:21}.) 
We further labeled their distinct Berry curvatures as $\Omega_{\ell ,\pm} ({\bs k})$, 
which satisfy $\Omega_{\ell ,\pm} ({\bs k}) = k^{-2}  \Omega_{\ell,\pm}( \theta,\varphi )$ and 
$\Omega_{\ell,-} (\theta,\varphi ) = - \Omega_{\ell,+} ( \theta,\varphi )$. 
In the third step, we have used 
$n_f (E_{\ell,\pm} ({\bs k})  )\approx n_f ( \epsilon_\ell ({\bs k}) ) \mp | \vec{f}_\ell ({\bs k}) | 
~   \delta ( \epsilon_\ell ({\bs k}) - \mu) 
~ ~\text{for  } W k  \gg V^2 / \Delta \epsilon $. 
In the last step, we have assumed the spherical Fermi surfaces for both $\epsilon_{1(2)}$ with the wave vectors $k_{F}$ to separate the radial and angle integration,  
$
\delta ( \epsilon_ {1(2)} ( k ,\theta, \varphi) - \mu) \approx  \frac{1}{W a^2  } \delta ( k ^2 - k_{F}^2 )
=  \frac{1}{2 k_{F}}  \frac{1}{W a^2  } [ \delta (k + k_{F}) + \delta ( k - k_{F}) ]$ 
and this leads to 
$ \int_0^\infty dk ~ (k a) ~ \delta ( \epsilon_ {1(2)} ( k ,\theta, \varphi) - \mu) 
=
 1/(2a W) 
$
where $ka$ in the integration is from $f_\ell (k , \theta , \varphi) = (ka) f_\ell(\theta, \varphi) $.
\\

Within the spherical Fermi surface approximation, we are left with the angle integration and realize that the integration of Berry curvature is not proportional to the Fermi wave vector $k_F$ in the limit of $W k  \gg V^2/  \Delta \epsilon$. 
Thus, the band split from the original doublet because of the perturbative quadrupolar order   
leads to the Berry curvatures, independent to the magnitude of Fermi wave vector but proportional to $V^2/ (\Delta \epsilon ~W) $ only with azimuthal and polar angle dependence that captures broken inversion and mirror symmetries.

\end{document}